# Time of Flight Mass Spectrometry with Direct Extraction of a Uranium Plasma

James O. F. Thompson, S. Tahereh Alavi, Justin R. Walensky and Arthur G. Suits*

Department of Chemistry, University of Missouri, Columbia, Missouri, 65211, United States

**Abstract**

We demonstrate a Wiley-McClaren Time of Flight Mass Spectrometer modified to allow direct extraction of ions from a laser-induced plasma. Unlike many other methods that utilize mass spectrometry to investigate laser-induced plasmas, no collisional cooling or further ionization of the plasma is required prior to acceleration, allowing measurements of the plasma ion composition to be obtained directly. Furthermore, we show using the laser ablation of gadolinium as an example that we are able to obtain the translational energy distribution of the plasma directly and infer information about the relative composition of ions within different regions of the plasma plume. The approach is then applied to laser ablation of a uranium sample as a step toward probing the chemistry under conditions relevant to a nuclear fireball for nuclear forensics applications.

*suitsa@missouri.edu

**Graphical Abstract**

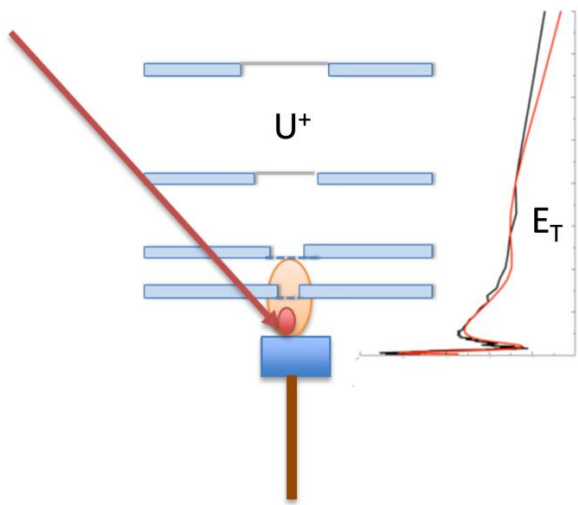

## 1. Introduction

Over the last few decades, the properties of plasmas produced by laser ablation processes have been studied extensively. One of the main goals of these studies has been to develop laser ablation methods and laser induced plasma spectroscopy as a tool for elemental analysis.[1-17] An important application of such studies in which laser ablation based analytical techniques have been used is nuclear forensic investigations.[18, 19] Laser ablation based mass spectrometry techniques are among the chief analytical approaches used in nuclear forensic investigations because they can provide highly accurate elemental and isotopic information while requiring less mass compared to alternative analytical methods available for such studies.[20] However, direct characterization of actinide laser ablation plasmas has rarely been pursued, yet these can provide conditions approaching those of a nuclear fireball. Such studies thus offer a means of investigating reactions and condensation of materials relevant to nuclear forensics investigations. We report our initial investigations along these lines here.

Given that the internal temperatures of a plasma can be on the order of hundreds to many thousands of kelvins,[21-23] understanding the dynamics of such systems can be challenging. Additionally, laser induced plasma typically have a non-Boltzmann energy distribution adding further complexity to any modeling of the plasma system.[24] Even though laser ablation is used in a number of fields, there are still significant gaps in the understanding of how these ablation events progress.

Plasma properties are typically experimentally studied either through optical emission spectroscopy or mass spectrometry methods. In optical emission spectroscopy methods, the produced plasma is studied directly during its ignition and initial expansion by recording the light emitted from the plasma which provides a fingerprint for both the species in the spectra and electronic temperature, while in mass spectrometry methods the plasma is studied using debris or soot collected from the plasma, providing insight into the species ultimately formed after the plasma has cooled. Optical emission studies of the laser induced plasma are referred to as laser-induced breakdown spectroscopy (LIBS).[25] There are a number of examples of this technique used to study the ablation of transition metals,[26] inorganic materials[27] and alloys.[28, 29] Although the focus is generally on atomic species, recently, Russo and co-workers have utilized LIBS to identify many spectral lines of diatomic molecules.[30] Through this, they were able to use the isotopic shift of these emission lines to characterize the abundance of different isotopes present in the plasma. However, this is only possible in plasma with a well-characterized target and a small number of elements present due to the inevitable spectral congestion that would occur in a complex plume when detecting molecular emission. In the mass spectrometry technique, post ablation debris of the plasma is studied to understand its composition. Typically, this involves collecting the particulates from the plasma into a free jet or a molecular flow, cooling the particles in the process, and transporting the material to a mass spectrometer.[31] The most commonly used approach is laser ablation inductively coupled plasma mass spectrometer (LA-ICP-MS).[4, 31-37] In comparing the two methods briefly, the principal difference is that such mass spectrometry-based methods do not provide any direct diagnostic information regarding the initial plasma conditions, although they can give efficient elemental composition and isotopic ratio information. In contrast to this, optical emission-based methods can identify the conditions of the plasma. Furthermore, in nuclear forensic studies, using a portable LIBS unit is more time efficient and less expensive than a mass spectrometric approach [38]. However, such methods have lower detection sensitivity compared to mass spectrometry-based methods and show difficulty in identifying isotopic information of the species inside the plasma. The other significant drawback of LIBS methods is that they are limited to identifying the composition at early part of the plasma (around a few hundred ns), while the dynamics of the plasma evolution continues over a much longer period of time.

To overcome this, typically modern instruments combine both of these techniques together and work in tandem[39], which is extremely successful given the non-invasive nature of the emission spectroscopy detection. Nevertheless, there are some issues with this dual functionality set-up. Firstly, there is a cost to setting up a dual function spectrometer such as the iCCD cameras required for the capture of the LIBS emission and the vacuum system required for the mass spectrometer. Secondly, where the interest is in hot plasmas, conventional explosives or indeed a nuclear fireball, line broadening happens and results in a loss of information due to the overlapping of spectral features.

In this work, we present an alternative approach using a modified Wiley-McLaren time of flight mass spectrometer [40, 41] to directly extract ions formed in the plasma during the ablation event. This setup is capable of overcoming many of the aforementioned issues whilst maintaining the principal advantage of LA-ICP-MS which is observing dynamics in the plume over a much longer time domain. With this instrument we are able to probe the plasma directly and extract translational energy distributions of the plume alongside its ion composition.

There are a number of challenges in attempting to extract ions directly from a plasma in this way. The biggest issue is that the plasma is so dense that high voltage extraction of ions causes massive space charge effects. This leads to a reduction in mass resolution of the spectrometer and can lead to potential damage of the ion detector. Also, as there is no resistance to the plasma expansion, the plasma maintains extremely high initial particle velocity which can blur the mass spectra. These technical challenges must be overcome to facilitate accurate ion detection, and the solutions we employed are discussed here.

In this work, we aim to use the aforementioned laser ablation-based mass spectrometry method to produce high temperature uranium (U) plasma in order to gain insight into the chemistry of a nuclear fireball. Characterization of a U plasma using high laser power can provide useful nuclear forensic information: Isotope ratio characterization, and quantitative and qualitative analysis of radioactive samples. These signatures have been studied using LA-ICP-MS and LIBS techniques [42-46], however, these approaches do not probe the conditions of the expanding plume directly.[11] In the present study, we extract the plume through small apertures to gain direct access to the conditions in the plasma. Although this approach cannot yield mass resolution adequate for isotope ratio analysis under high power ablation, it is complementary to the more established techniques in gaining direct access to ion charge state and translational energy distributions. After calibration of the mass spectrometer and investigations of its performance using Al and Gd targets, the same experiment with U is performed to begin investigation of a hot actinide plasma.

**Experimental**

For these studies we use a Nd:YAG laser (Quantra-Ray DCR 3, Spectra Physics) which provides up to 1 J/pulse (10 Hz repetition rate (RR)) at 1064 nm with a pulse duration of 8 ns. The laser has an optional frequency conversion unit to provide the 2$^{nd}$ and 3$^{rd}$ harmonics of the fundamental (532 and 355 nm respectively). The laser beam is then sent to a set of folding mirrors into a periscope to raise the beam above the ion optic assembly within the apparatus. The apparatus consists of a reaction chamber, a source chamber, a time-of-flight (TOF) drift tube and finally the detector. Both chambers are pumped using turbomolecular pumps (Varian,T-551) and (Osaka,TG1113MCW). All experiments were performed under vacuum at a base pressure of 10$^{-7}$ torr. The interaction chamber of the apparatus is the reaction and ion optics region where the plasma is formed and accelerated into the TOF drift tube region using a modified Wiley-McLaren accelerator composed of four electrodes shown in Figure 1. The distances between the electrodes are the same as those described by Townsend and coworkers [47] for direct current sliced velocity mapped ion imaging (employed in other experiments) and are noted in the figure. The flight time of the generated ions can be acquired using Wiley-McLaren equations shown below.

$$U = U_o + qsE_s + qdE_d \tag{1}$$

$$T = T_s + T_d + T_l \tag{2}$$

$$T_s = \frac{\sqrt{2m(U_o+qE_s)}}{qE_s} \tag{3}$$

$$T_d = \frac{\sqrt{2m}}{qE_d}\left[\sqrt{U} - (U_o + qsE_s + qdE_d)\right] \tag{4}$$

$$T_l = L\sqrt{\frac{m}{2qU}} \tag{5}$$

Here, $U_o$ and $U$ are the initial translational energy and the post acceleration total translation energy of the particle, $q$ is the elementary charge, $s$ is the distance from the position where the ions are extracted to the extractor electrode, $d$ is the distance between the extractor electrode and the grounded electrode and $L$ is the length of the field free region of the flight tube. $E_s$ is the strength of the electric field at position $s$ and $E_d$ is the potential of the extractor electrode. Finally, $T_s$, $T_d$, $T_l$ and $T$ are the times the particle spends traveling from the acceleration region to the extractor electrode, from the extractor electrode to the grounded electrode, from the grounded electrode to the detector, and the total flight time, respectively. The electrodes were constructed of a mirror-polished stainless steel and a thin nickel mesh (Precision Eforming, MN12), 45 lines per inch with 88% transmission) used to cover the orifice of each electrode except for the repeller. The plasma density is reduced in two steps from the ablation region to the acceleration region. As shown in Figure 1, the first electrode features a single aperture (of 75 ± 20 μm diameter) in the center of the plate that reduces the initial density. Taking inspiration from Ogawa and co-workers,[48] a "pepper pot" style grid consisting of 9 holes (3x3, 1 mm spacing, 50 ± 20 μm diameter each) was fashioned on a 0.004" thick plate of stainless steel mounted on the repeller. The diameter of each hole was confirmed by optical microscopy. During the acceleration, the middle two electrodes of the assembly (repeller and extractor electrodes) are pulsed to 1500 V and 1220 V respectively from ground while the other electrodes are grounded for the entire experiment. The electrodes were pulsed using a high voltage pulser (DEI, PVX-4140, rise time 100 ns), for 50 μs to provide a constant energy impulse to accelerate the ions. Additionally, resistors of 150 and 50 Ohms were placed inline from the pulser to these two electrodes.

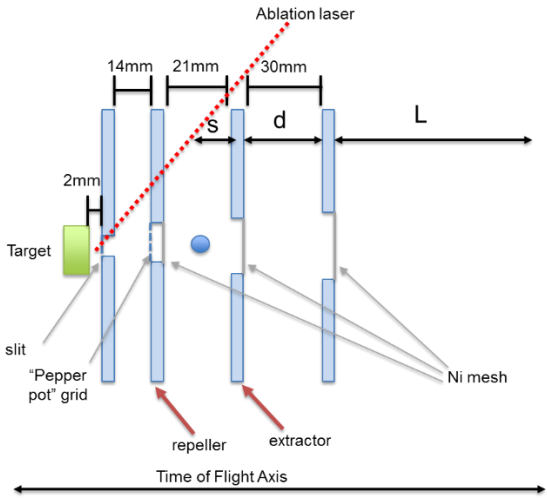

Figure 1 – Schematic of Wiley McClaren Accelerator. The red line indicates the path of the ablation laser. The blue circle indicates the center of the accelerator volume.

For the target, a 25mm x 25mm x 1 mm gadolinium sheet (693723 , Sigma-Aldrich), aluminum foil (99.9%) and natural uranium plates (425-349-30, Goodfellow) were used. No additional cleaning or polishing of the surfaces was performed aside from that of the ablation itself as discussed for the uranium surface below. In each experiment the target plate was mounted in the PEEK (Polyether Ether Ketone) assembly onto a motorized 2D translational stage within 2 mm of the entrance electrode of the accelerator located in the reaction chamber. The light was focused onto the target using a 30 cm AR (Anti Reflection) coated fused silica lens through a pair of externally mounted irises prior to entering the chamber through a 1″ diameter 2 mm thick fused silica window. To allow the light to pass through the ion optics, 0.5 cm diameter holes were cut into the electrodes at an angle of 27° to the electrode surfaces. Based on our focusing conditions, we estimate the highest power density of our 8 ns pulsed laser beam to be ~5 x $10^{11}$ W/cm$^2$ at the target, assuming a diffraction limited beam spot at the focus. Due to the geometry of the experiment, the laser does not produce ions directly in the accelerator region, therefore any ions detected must be formed either during the ignition event in the laser focus or from collisions with the plasma as it expands. They also can result from electron impact ionization during the initial acceleration as the electrons are torn from the neutral plasma.

Once produced, the plasma travels from the target into the assembly in a field-free condition. After an allotted time has passed, termed the extraction delay, the accelerator is pulsed to accelerate the ions into the drift tube. The ions were ultimately detected using an 40 mm dual microchannel plate (MCP) imaging detector (Beam Imaging Solutions) with the front plate grounded and the back plate biased to 1900-2200 V. The time-of-flight was recorded directly from the MCP using a home-built signal decoupler fed to a 250 MS/s digitizer (NI, PCI-5114). The data acquisition was controlled using the NI Scope software on LabView 2017. The mass spectrum was then recorded and averaged typically for 50 laser shots at each delay. Furthermore, in order to investigate the effect of the field-free propagation time, the delay of the HV switch relative to the laser Q-Switch was controlled by a delay generator (BNC). The change of the delay and acquisition of the mass spectrum was then automated using an in-house built LabView program.

## 2. Accelerator Performance

To determine the overall performance of the spectrometer we performed laser ablation experiments on Al using 1064 nm at 30 mJ/pulse. Shown in Figure 2 is the TOF spectrum of the Al target at an extraction delay of 6 µs. The mass spectrum has been cropped to show only the singly charged $Al^+$ peak, however, the doubly charged $Al^{2+}$ peak is present and found at 5.8 µs. The width of the $Al^+$ peak is larger than what one would expect for a Wiley-McLaren spectrometer for a single isotope mass spectrum. The broadening occurs due to the velocity dispersion that ions initially have prior to acceleration. The origin of this width can be understood by considering the dispersion of the initial translational energy of the particles within the accelerator acquired from the ablation process, and the range of velocities of the ion packet formed over the full length of the acceleration region. To model this, we reproduced our accelerator using SIMION 8.2.[49] In these simulations, we recorded the arrival time of a number of Al ions starting from 22 different positions within the accelerating volume in 1 mm increments between the 2nd and 3rd electrodes. The ions at each position were given a distribution of initial translational energy in the z-direction (along the TOF axis) centered on the energy required to reach their starting position within the accelerator in 6 µs. This consisted of 20 ions with a Gaussian distribution of initial translational energy centered on the energy required for the ions to arrive at the specified point in the accelerator in 6 µs. The FWHM of this distribution was also equal to the same translational energy in order to model dispersion of ion energies within the accelerator. These (440 in total) ion trajectories were recorded and counted in bins of 20 ns. This built up a TOF profile of the Al ions in the plasma within the accelerator. A 5-point moving average smoothing was applied to account for the 100 ns rise time of our accelerator field ramping. The simulation, shown as the red curve in Figure 2, is in fair agreement with the experimental data.

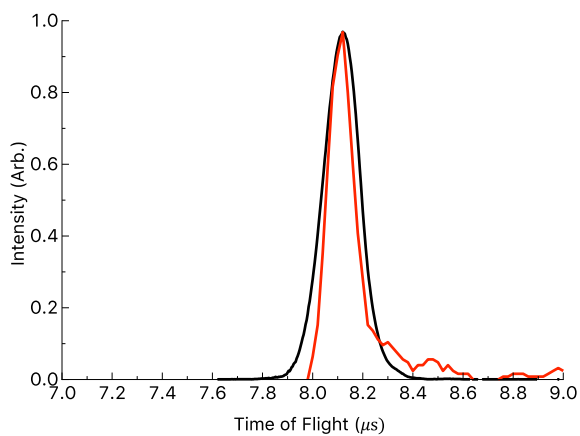

Figure 2 – Time of Flight mass spectrum of Al. The spectrum shown in black is the experimental spectrum and the spectrum shown in red is that produced from our SIMION simulations.

## 3. Gadolinium Ablation Results

In order to demonstrate the abilities of the apparatus and characterize the time of flight mass spectrometer, we chose gadolinium (Gd) metal as the target. Gd was chosen due to its large number of isotopes at a high mass-to-charge ratio (around m/z 157). This metal has also been studied before using a combination of direct plasma extraction mass spectrometry and laser induced breakdown spectroscopy at vastly lower ablation energies than employed here. [50] In work by Song and co-workers, a Gd plate was mounted directly onto the accelerator of the time-of-flight mass spectrometer, in a manner similar that of most MALDI (Matrix-assisted laser desorption/ionization) type instruments. They demonstrated the ability to maintain isotopic resolution at this high m/z under very low power of laser ablation and DC extraction of ions. However, in their study, the mass resolution was compromised even at moderate laser powers and they did not investigate the plasma conditions at high power.

Before recording the gadolinium TOF mass spectrum we investigated the mass loss of Gd via laser ablation at our high-power condition (30mJ/pulse) to estimate the particle density within our accelerator. To do so, the sample was ablated a number of times for an hour and the total mass removed was recorded. The mass loss was determined to be 63.1 ± 8.6 ng per laser shot. This is on the same order as that reported by Russo and coworkers for a copper surface under similar ablation conditions, except that plume expanded against gases at atmospheric pressure. However, they noted the volume of material ejected was found to be independent of gas for He, Ne and Ar, so we do not expect this to be different for expansion into the vacuum.[54] Our determination corresponds to $2.4 \times 10^{14}$ atoms (or 0.4 nmol) ablated in a single shot. Assuming a hemispherical expansion, and that all the removed mass went into the initial plume, we calculate that the total density of the plasma before the first slit is $\sim 1.4 \times 10^{16}$ atoms/cm$^3$. We estimate that the total amount of material that passes through the first slit is approximately $4 \times 10^{10}$ atoms per laser shot (about a $10^4$ reduction in total number). Furthermore, we estimate that through the second slit (the pepper pot grid) the number of atoms is further reduced another factor of $10^3$ to $4 \times 10^7$. At these reduced densities we do not observe any Coulomb repulsion when extracting the ions from the plasma using the accelerator.

To characterize our modified TOF mass spectrometer we performed a laser ablation experiment on a Gd sample using 1064 nm, 30 mJ/pulse. Figure 3 shows the extraction delay-dependent TOF mass spectrum of the Gd plasma. The spectrum was recorded at delays from 0 to 60 μs in 0.5 μs steps. In addition to the two main expected signals at 158 m/z and 79 m/z of $Gd^+$ and $Gd^{2+}$ there are a number of additional peaks in the data that vary as a function of extraction delay. These additional signals are identified as background gases ionized in the chamber by plasma species. This is most likely through either collisional ionization from charged species within the plume or from electron impact ionization during the acceleration process (with a maximum available energy around 300 eV). These signals generally appear at delays consistent with the thermalized region of the plasma except for m/z=18 (water) and 40, which persist for much longer times. We assign m/z=40 to argon, which is used in the production of Gd. This mass was also seen by Song et al. in their Gd ablation and curiously assigned to potassium without comment.[51] It seems that it is trapped in the foil giving signals that persist to much longer times than other background signals such as m/z=28. It is interesting that the water signals also show a much longer appearance profile much like argon, perhaps suggesting some is also trapped in the sample.

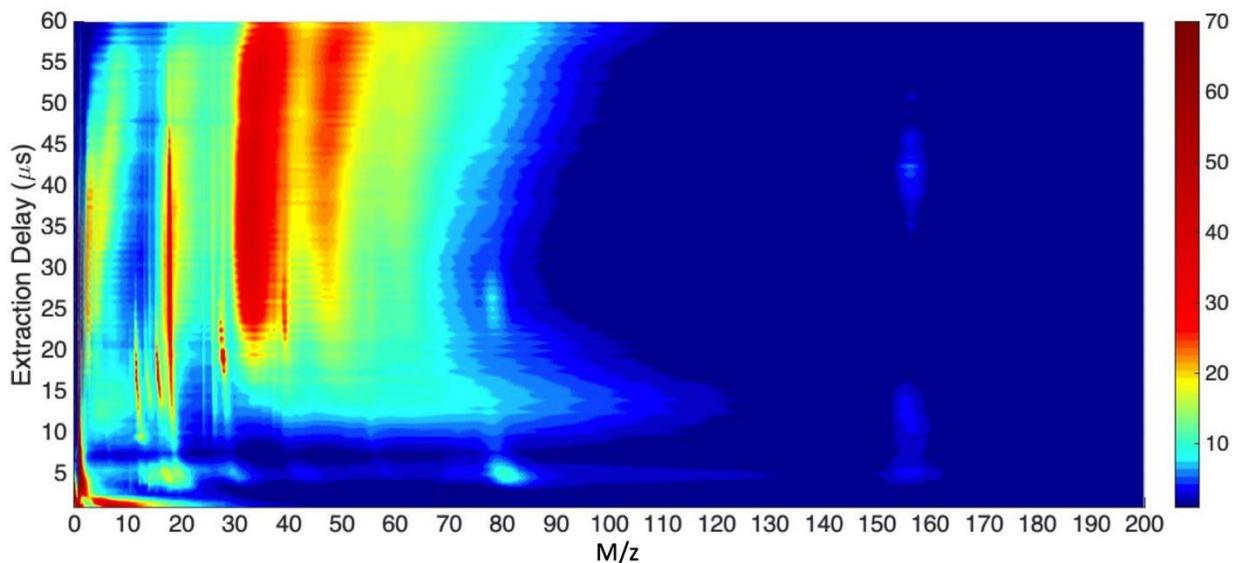

Figure 3 – Extraction delay-dependent time-of-flight mass spectrum of gadolinium laser ablation plasma.

The data shows that prior to acceleration in the mass spectrometer, the Gd ions have three distinct arrival times centered at 5 µs, 12 µs and 42 µs, indicating a distribution of ion velocities within the plasma plume. We suggest that these three regions correspond to those ions found within the hot initial shockwave layer of the plasma (5 µs) which may be followed by a reflected shock from the back of the first electrode and some attenuation. Finally, there is the colder thermalized part of the plasma (42 µs). This cooling we suggest comes mainly from collisions within the plume itself behind the first aperture. The identification of the regions as either shockwave or partially thermalized is further supported by the data. The early time region (around 5 µs) has an intense non-Boltzmann distribution of $Gd^+$ and $Gd^{2+}$ ions that would indicate a region of the plume that is extremely hot/energetic. It also has very high translational energy. The late time region identified as the thermalized core of the plume is supported by comparing the average translational kinetic energy of the $Gd^+$ to that of the background and will be discussed later in this section. Based on our initial measurement of the accelerator performance using Al, we would expect a significant reduction in the width of the $Gd^+$ peak as a function of extraction delay as the colder parts of the plume have lower translational energy dispersion.  Figure 4 shows exactly that.  Figure 4a shows the Gd$^+$ channel at 5 µs whilst Figure 4b depicts the TOF signal at 45 µs.  The data does not show any clear indication of the various isotopes of Gd, $^{154}$Gd (2.18%), $^{155}$Gd (14.80%), $^{156}$Gd (20.47%), $^{157}$Gd (15.65%), $^{158}$Gd (24.84%), and $^{160}$Gd (21.86%), but a small shoulder in the Figure 4b spectrum is from $^{160}$Gd.  Past laser ablation studies of Gd have been able to identify these isotopes but with a different configuration mass spectrometer at significantly lower power, 1.2 MW/cm$^2$ previously compared to our 500 GW/cm$^2$.  Once again, however, the simulations of the Gd$^+$ TOF spectrum are in fair agreement with what we record.  The simulations were run in the same manner as for Al described above, but in this case, we ran the simulation for each isotope separately and scaled the results by their abundance within the sample.  Therefore, we infer that the translational energy distribution within the Gd$^+$ at 5 µs and 45 µs are 40.6 eV, and 0.5 eV, respectively, roughly matching the spread of the translational energy of the particles within the accelerator based on the arrival time.  The simulations do suggest that we should be able to resolve the $^{160}$Gd isotope partially at 45 us extraction delay, which appears as a shoulder in the experimental data.

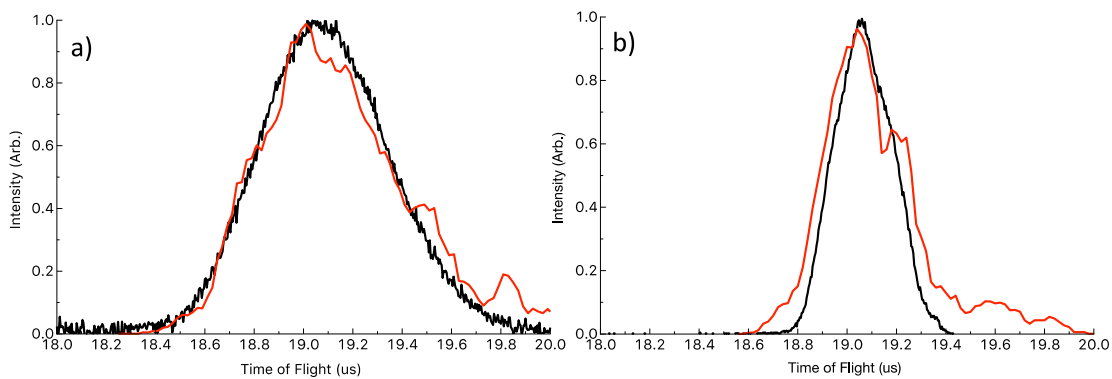

Figure 4– Time of Flight mass spectra of $Gd^+$ at delays of 5 µs (a) and 45 µs (b) between laser ablation and ion extraction. The spectrum in black shows the experimental TOF and the simulation is in red.

In addition to determining the spread of the translational energy within the accelerator using the width of the mass spectrum, the data in Figure 3 can also be used to determine the translational energy distribution of the plasma directly, by measuring the Gd$^+$ signal as a function of extraction delay.  As we know the path length of the plasma prior to acceleration, we can calculate the velocity required for the ions to arrive in the accelerator at different times; this is in fact the same basis used for determining the translational energy distribution given to the ions in different regions in the accelerator in our SIMION simulations above.  For Gd$^+$, the translational energy distribution obtained based directly on the field free flight times (i.e., extraction delay), and that based on the peak width of the mass spectra, are in agreement. In these simulations, we use a single ion position within the accelerator, the one at which the focusing of the ions onto the detector is optimized for our applied potentials, as our position to calculate the subsequent time-of-flight of the $Gd^+$ ions using the Wiley-McLaren equations. The use of a single ion position is both to simplify the analysis of the SIMION trajectories and also to focus on the ions which will be readily detected by our apparatus. This is due to a) ions accelerated in regions away from the central volume will either not focus as tightly and been seen as a structureless background to the mass spectrum and b) due to geometric constraints of the electrodes ions are unlikely to have much transverse momentum to take them away from the central axis of the spectrometer. Additionally, in order to correctly fit the experimental data, we must apply a scaling factor of 0.75 to the calculated TOF. This correction may be due to plasma screening the full effective field of the accelerator, leading to a longer ion time of flight (and contributing to some additional blurring of the TOF as mentioned above). The results of this simulation are shown in Figure 5. Figure 5a shows the experimental data, 5b shows the simulation and Figure 5c shows the translational energy distribution of the $Gd^+$ over the entire plume expansion. The simulation shown in Figure 5b was produced by modeling the arrival time distribution in the field-free TOF axis with three Gaussians to correctly capture the intensity distribution of the particles. One of the key features of the simulation is a predicted curvature in the arrival time of the $Gd^+$ particles in the early time window. This bend arises from the fact that for the $Gd^+$ ions to arrive in the accelerator within such a short time, the initial translational energy of the particle $U_o$ must be comparable to the total translational energy of the particles after acceleration. This curvature in fact, is a direct measure of the change in translational energy as a function of the *field-free* plasma flight time. Using this, a translational energy spectrum of the Gd$^+$ particles within the plasma is obtained and shown in Figure 5c. A scaling factor of $t^3$ was incorporated to correct for the Jacobian in the transformation of the intensity as a function of time to intensity as a function of translational energy. The translational energy distribution spectrum shows two main peaks. These are located around 0.55 eV and 6 eV, but the latter extends to well beyond 50 eV.

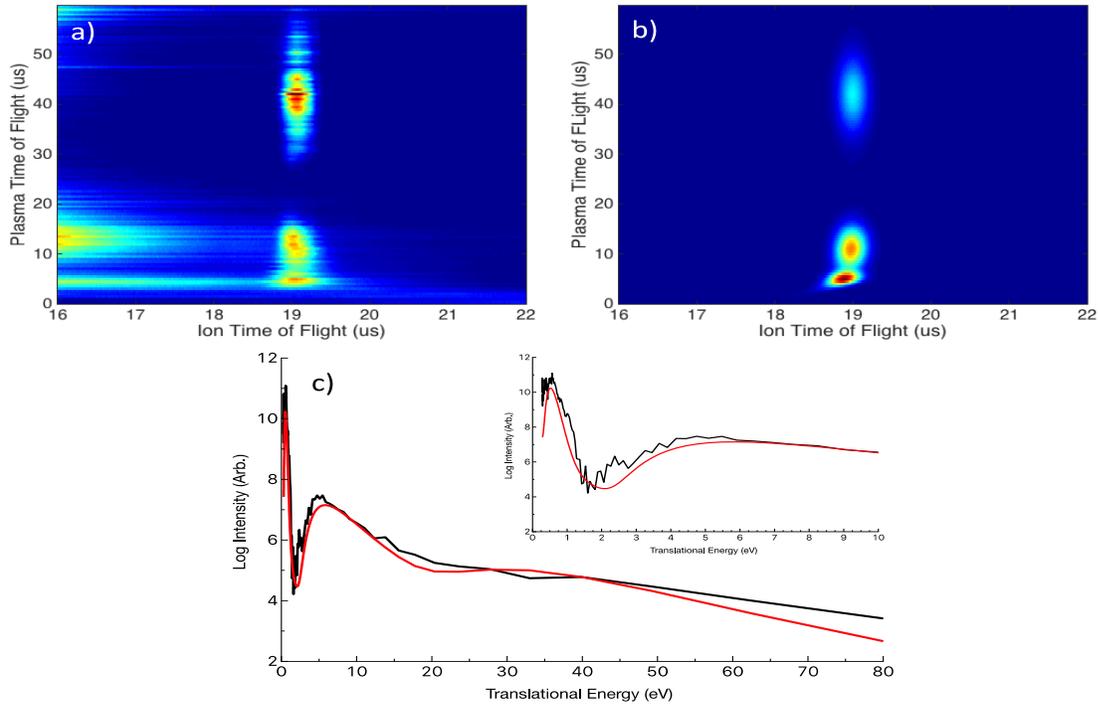

Figure 5 – a) Time of flight mass spectrum of the Gd$^+$ channel acquired at 30 mJ/pulse using a 1064 nm laser source. b) Simulation of this data using the Wiley-McLaren equations. c) Translational energy distribution of the Gd$^+$ channel over all extraction delays with the experimental distribution in black and the simulation in red. The inset in (c) shows the same distribution and fit between 0 and 10 eV.

After converting from energy to temperature, assuming this energy is placed in only a single translational degree of freedom, these regions correspond to translational temperatures of $1.2 \times 10^4$ K, $1.4 \times 10^5$ K and $8.1 \times 10^5$ K. Russo and coworkers simulated ablation of a copper surface against argon at atmospheric pressure under very similar laser conditions to those reported here. They obtained peak temperatures at the contact surface of the plume with the background gas of $3 \times 10^5$ K, on the same order as that observed here. However, they found this was somewhat in excess of experimental measurements, and assigned the discrepancy to omission of energy dissipating processes such as ionization.[52, 53]

It is important to note that we do not see evidence of ion cluster formation in any of our measurements. We believe that this is due to the fact that the plasma is too hot and no such stable clusters are being formed under the conditions of our study. However, we have not employed any post ionization methods (such as an additional laser pulse to further ionize the plasma) in our current data, so we cannot infer any information about neutral cluster formation within the plasma that could be occurring as the plasma cools.

4. **Uranium Ablation Results**

We now turn to the uranium plasma, which is the focus of these studies. To characterize the uranium plasma, we perform a laser ablation experiment with two different wavelengths, 1064 and 355 nm, and at different laser powers (3,10 and 30 mJ/pulse). Figure 6 shows the extraction delay-dependent TOF mass spectra of uranium ions under these different ablation conditions. In this case the uranium surface was cleaned by prior ablation and free of oxides. The two main features in all laser powers and wavelengths belong to $U^+$ and $U^{2+}$ channels. Just as is the case in the Gd data, we see no sign of cluster formation. The TOF mass spectra obtained at different conditions show that the plume composition changes as a function of laser power and wavelength. At higher power the uranium signal becomes bimodal with ions arriving both at short and long extraction time delays. In addition, we see an increase in background signals with laser power. The production of doubly charged uranium ions as well as background signals also increases at the shorter wavelength. We did not observe any evidence of separation of the two uranium isotopes because of the wide range of initial translational energies that the accelerator compresses as discussed further below.

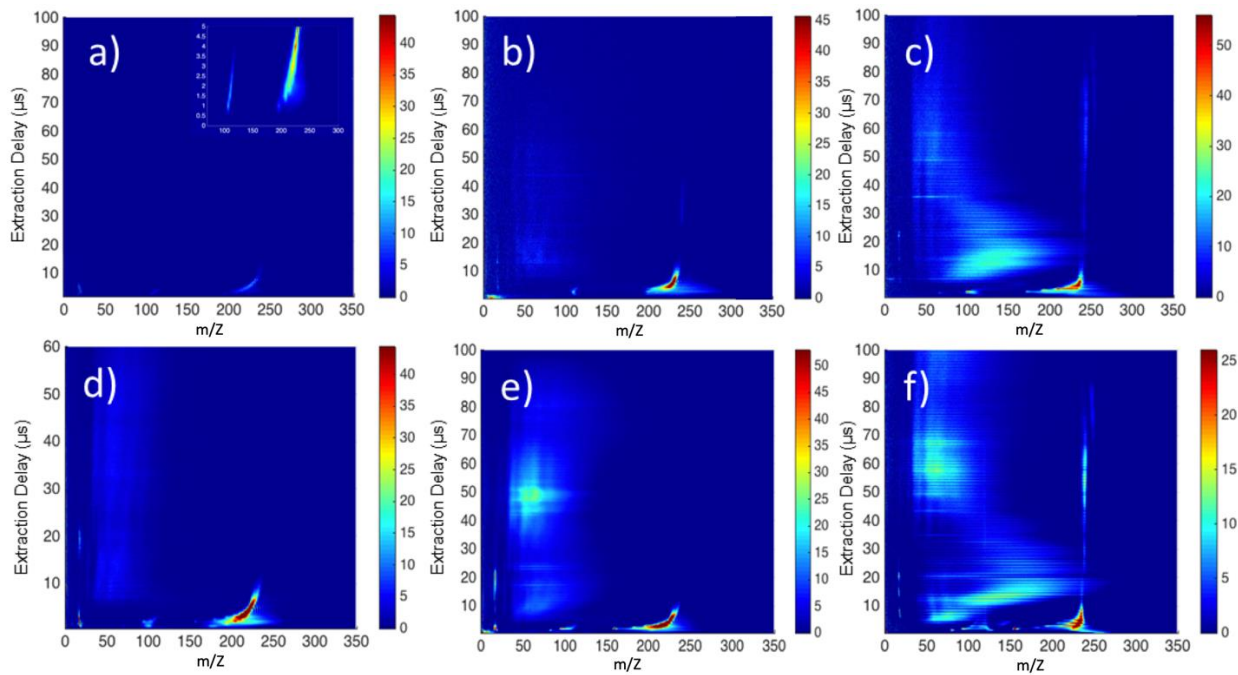

Figure 6 - Extraction delay-dependent TOF mass spectra of uranium plasma. The 1064 nm ablation at 3, 10 and 30 mJ per pulse are shown in panels a) – c) and the 355 nm ablation also at 3, 10 and 30 mJ per pulse are shown in panels d) – f) respectively.

In all cases, the arrival time of the $U^+$ is the same if not slightly earlier than the $Gd^+$ suggesting that the uranium plasma formed in these conditions is a much hotter, more energetic plasma. In either case we can assign the peak at 4 µs and 55 µs as belonging to the initial shockwave and the thermalized part of the plume, respectively. The translational energy of these two components are calculated to be 90 eV and 0.5 eV. Again assuming energy partitioning into one translational degree of freedom this would indicate temperatures of $2.1 \times 10^6$ K and $1.1 \times 10^4$ K. Although the initial impulse peaks at 4 µs, it begins to appear as early as 2 µs, implying a kinetic energy that is much higher, up to 350 eV. The thermalized temperature agrees well with our calculation for the later component of the $Gd^+$ temperature. It is interesting to note that we do not see the contribution identified as a reflected shock in the U data. This could be due the reflected shock signal becoming merged with the initial shock and unresolvable, but more detailed modeling will be needed to confirm this.

On comparing the uranium plasma signal in the late extraction delay region (>20 µs) we observe that a slower $U^+$ signal develops and grows more intense with increasing laser power. In addition, the distribution of the ions grows broader, visible over a much larger range of extraction delays. We attribute this to an increase in the fraction of the ions retaining their charge after passing through the high collision region before the first aperture on the accelerator. This would suggest a higher ratio of ion to neutral species in the plasma at higher ablation energies, consistent with the general picture that higher ablation energy leads to a hotter and more excited plasma.

Figure 7 shows the U$^+$ mass peak obtained at different laser powers at 1064 and 355 nm and the early extraction delay of 6 µs. Additional smoothing of the data was done in the case of 3 mJ/pulse to remove a ringing artifact from the data. At each wavelength the data is compared with a simulation (red curve) performed by SIMION 8.2 as explained before. Here, to model the data, we consider a 15% dispersion of translational energy around each position in the accelerator, while that for modeling Gd this was much larger. This suggests that this dispersion is more closely related to the spread in ion velocity at each position rather than spread in initial translational energy. This figure shows that change in the laser beam intensity does not change the overall peak width of the uranium signal significantly, although there is a tail to shorter times (higher energy) that is especially apparent in the 1064 nm data. In addition, we see a narrower width for UV laser ablation compared to IR.

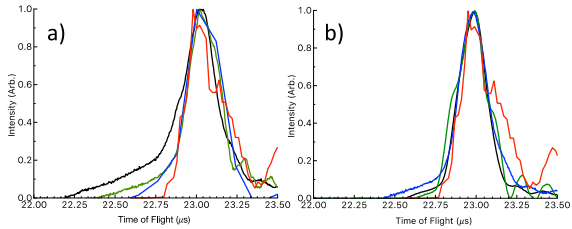

Figure 7. TOF mass spectrum of uranium plasma, ablated at 1064 nm (a) and 355 nm (b). In each plot we show results at three different laser intensities: 3 mJ/pulse (black), 10 mJ/pulse (green) and 30 mJ/pulse (blue). In each case the simulation is overlaid in red.

In the U⁺ channel we see a curvature at early times similar to the one observed in Gd⁺ data but much more pronounced. Again, we ascribe this curvature to the initial velocity of uranium ions in the TOF direction adding to the imparted momentum gained from the accelerator. The simulation was done using Wiley-McLaren equations as discussed above.

Figure 8 shows the experimental and simulated delay-dependent TOF mass spectra of the U⁺ channel along with its inferred translational energy distribution over the entire plume. The experimental spectrum is the same as that shown in Figure 7c but plotted in TOF rather than mass/charge, over the region of the U⁺ channel. Two Gaussians were used to model the intensity profile of the U⁺ channel centered at 90

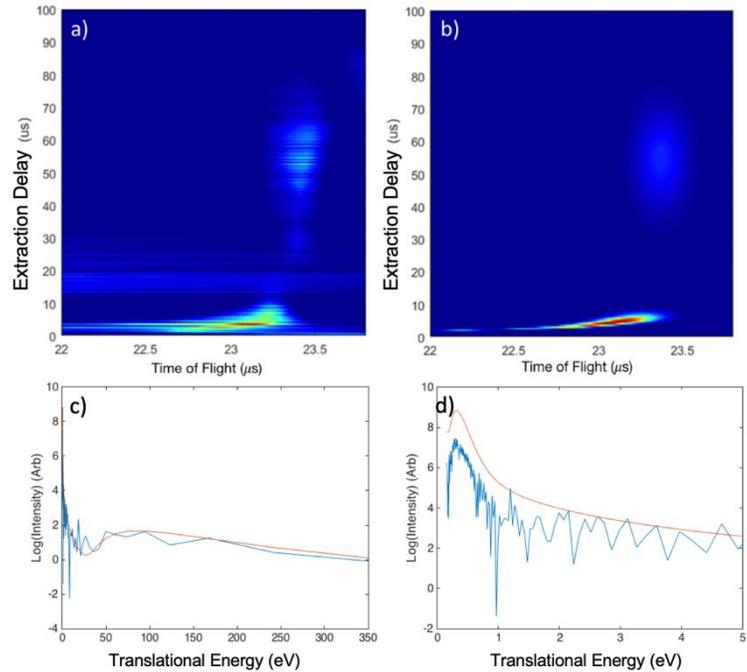

Figure 8 - a) Uranium mass spectra as a function of extraction delay b) Simulation of the data in (a). c)Translational energy distribution of U⁺ over the entire plume expansion with the raw data in blue and the fit in red. d) Expanded view of the distribution between 0 and 5 eV.

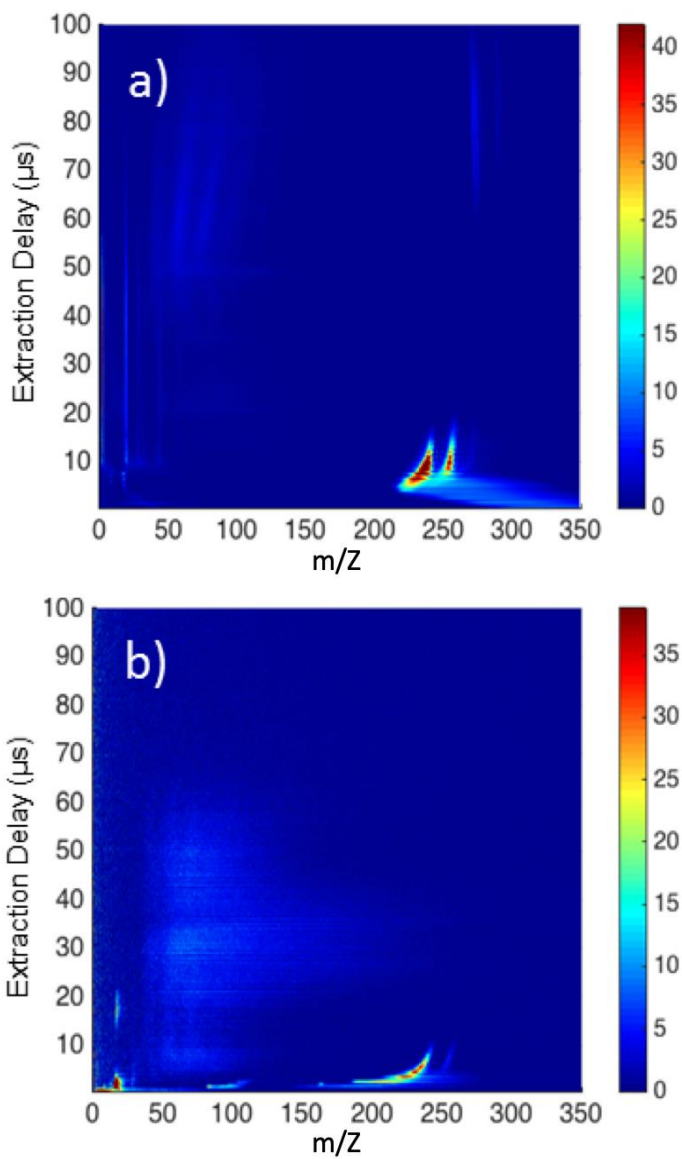

Figure 9 – Delay-dependent TOF mass spectrum of oxidized uranium metal. Ablation is performed at a) 1064 nm and b) 355nm. Laser power in both cases is 10

and 0.5 eV. The experimental data (Fig 8a) shows a very clear bend in the early plasma TOF, even more apparent than the $Gd^+$ in Figure 4, which is captured correctly by the simulation shown in Figure 8b. Figure 8c shows the translational energy distribution of the experiment and simulation. The overall trend of the distribution is very similar to the $Gd^+$ result.

Whilst we do not observe the formation of clusters in the uranium plasma, we are able to observe the effects of oxidation on the metal surface. Figure 9 shows the TOF mass spectra of an oxidized uranium surface ablated with a 10 mJ/pulse beam of 1064 and 355 nm (Figure 9a and 9b respectively). In both cases we observe that the $UO^+$ is the dominant uranium oxide species observed with a small amount of $UO_2^+$ also seen in the 1064 nm data. The oxidized target plasma otherwise is broadly consistent with observations of the cleaned uranium metal albeit with some minor differences. The 1064 nm data shows no formation of $U^{2+}$ at this laser intensity. In contrast, the 355 nm data does show the production of $U^{2+}$ but with a reduced amount of $UO^+$ present in the plume.

## 5. Conclusion

We have demonstrated a modified Wiley-McLaren Time of Flight mass spectrometer; designed to perform direct extraction of ions from an intense laser induced plasma at high ablation power. This approach demonstrates our ability to identify the effects of various experimental parameters on the plasma plume over a long period of propagation time. We examine the evolution of the plasma composition as a function of field-free time-of-flight of the laser plume and obtain a measure of the translational temperature of the plasma. We first characterize the performance of the system with aluminum and gadolinium targets combined with ion trajectory simulations, then apply these methods to study of an energetic uranium ablation plasma as a first step toward investigation of conditions approaching those of a nuclear fireball.

**Acknowledgments**

This work was supported by the Defense Threat Reduction Agency (DTRA) under project number J9BAP723008. The authors also wish to thank R. Scholtzhauer and C. Callais of the University of Missouri Physics and Astronomy workshop for assistance in fabricating parts of the apparatus.